\documentclass[preprint,12pt]{elsarticle}

\usepackage{amssymb}

\usepackage{amssymb}
\usepackage{amsmath}

\usepackage{mathrsfs}
\usepackage{multirow}
\usepackage{multicol}
\usepackage{todonotes}
\usepackage{hyperref}
\usepackage{subcaption}

\usepackage{float}

\usepackage{cleveref}
\usepackage{xcolor,colortbl}
\usepackage{booktabs}

\DeclareMathOperator{\offdiag}{offdiag}
\DeclareMathOperator{\diag}{diag}
\newcommand{\mo} {^{-1}}

\newcommand{\E} {\mathbb{E}}
\newcommand{\V} {\mathbb{V}}

\newcommand{\tr}{\mathrm{tr}}

\newcommand{\ignore}[1]{}

\newcounter{bla}

\journal{Computer Physics Communications}

\begin{document}

\begin{frontmatter}



\title{Using orthogonal projectors in multigrid multilevel Monte Carlo for trace estimation in lattice QCD}


\author[a]{A. Frommer}
\author[a]{J. Jimenez-Merchan\corref{author}}
\author[a]{F. Knechtli}
\author[a]{T. Korzec}
\author[b]{G. Ramirez-Hidalgo}

\cortext[author] {Corresponding author.\\\textit{E-mail address:} jimenezmerchan@uni-wuppertal.de}
\address[a]{School of Mathematics and Natural Sciences, Bergische Universität Wuppertal, 42097 Wuppertal, Germany}
\address[b]{J{\"u}lich Supercomputing Centre, Forschungszentrum J{\"u}lich GmbH, Wilhelm-Johnen-Straße 52428 Jülich, Germany}

\begin{abstract}
We introduce a multigrid multilevel Monte Carlo method for stochastic trace estimation in lattice QCD based on orthogonal projections. This formulation extends the previously proposed oblique decomposition and it is assessed on three representative problems: the connected pseudoscalar correlator, the trace of the full Dirac operator’s inverse $\tr(D^{-1})$, and disconnected fermion loops. For the connected correlator, variance reductions grow systematically with the time separation and lead to cost savings of up to a factor of 30 at large separations, outperforming both the plain Hutchinson’s estimator and the oblique formulation. For $\tr(D^{-1})$, reductions are more modest but remain systematic, with stronger effects on more ill-conditioned systems. Disconnected loops show no improvement, since their variance is dominated by local same-slice contributions not targeted by the decomposition.

\end{abstract}


\end{frontmatter}



\makeatletter
\def\ps@pprintTitle{%
  \let\@oddhead\@empty
  \let\@evenhead\@empty
  \let\@oddfoot\@empty
  \let\@evenfoot\@oddfoot
}
\makeatother

\section{Introduction} \label{sec:introduction}

Computing the trace $\tr(A)$ of a square matrix $A$ can be a challenging problem if $A$ is not directly accessible as is the case, for example, if $A$ is the inverse $A = B^{-1}$ of a given matrix $B$ or, more generally, a matrix function $A = f(B)$ of $B$.
Traces of matrix inverses arise in various applications, including data analysis \cite{barthelme2019estimating}, quantum information processing \cite{Fitzsimons2018}, and the study of fractals \cite{PhysRevLett.67.2974}. In this work, we focus on its relevance in lattice quantum chromodynamics (QCD), an approach used in theoretical physics which allows one to obtain observables related to the strong interaction between quarks and gluons nonperturbatively.

QCD describes the quarks and gluons as the elementary constituents of protons, neutrons, and other hadrons. The quark propagator is obtained by solving Dirac's equation. In lattice QCD, the Dirac equation is discretized on a 4-dimensional space-time lattice, where twelve degrees of freedom correspond to each lattice site: one for each possible combination of four spin and three color indices. 

Today, typical lattice sizes are of order 100 in each dimension, yielding matrices $D$ that represent the discretized Dirac operator with a dimension of the order of $10^9$.  Traces of the inverse of $D$ arise in several contexts---some of which we will explain in more detail in Section~\ref{subsec:physical_applications} below---and, since $D$ is large and sparse, a direct computation of $\tr(D^{-1})$ is impractical.

Hutchinson's method \cite{Hutchinson1990} is a widely employed stochastic approach used to compute an {\em estimate} $\hat{ \tr}(A)$ for the trace $\tr(A)$ of a matrix $A\in \mathbb{C} ^{n \times n}$:  
\begin{equation}
\label{eq:Hutchinson}
    \tr(A) \approx \hat{\tr}(A) = \frac{1}{N}\sum_{k=1}^N (x^{(k)})^{\dag}Ax^{(k)},
\end{equation}
\\
where $x^{(k)}$ are stochastically independent samples of random vectors $x$ in $\mathbb{C}^n$, having independent, identically distributed entries $x_i$ following distributions where $\E[x_i]=0$ and $\E[\bar{x_i}x_j]= \delta_{ij}$. If, for example, one takes the $\mathbb{Z}_4$ distribution,
\begin{equation}
\label{eq:z4}
    x_i \in \{-1,1, -i, i\} \quad \text{with equal probability } \tfrac{1}{4},
\end{equation}
then the variance of $\hat{\tr}(A)$ is 
\begin{equation} 
\label{eq:variance-Frobenius-Relation}
    \V[\hat{\tr}(D)] = \frac{1}{N}\|\offdiag(A)\|_F^2,
\end{equation}
where $\|\cdot\|_F$ denotes the Frobenius norm; cf.\ \cite{Bernardson1994}. We obtain the same variance if we take the uniform distribution on the unit circle. For other distributions we might have different expressions for the variance like $\tfrac{1}{2N}\| \offdiag(A+A^\top)\|_F^2$ for the $\mathbb{Z}_2$ distribution and $\tfrac{1}{N}\| A+A^\top \|_F^2$ for the standard normal distribution $\mathcal{N}(0,1)$; see, e.g., \cite{FROMMER2021}.

The trace estimator $\hat{ \tr}(A)$ is unbiased, satisfying $\E[\hat{\tr}(A)] = \tr(A)$. Its variance decreases proportionally to $\frac{1}{N}$, meaning that the number of samples $N$ must grow quadratically with the desired precision. Consequently, achieving higher precision entails a rapidly increasing computational cost for Monte Carlo estimations of this kind. 
For this reason, variance reduction techniques have become a major topic in trace computation for lattice QCD, and we will shortly review existing techniques in Section~\ref{sec:variance-reduction-techniques} before introducing a new variant relying on cost-efficient high-rank orthogonal projections in Section~\ref{sec:orthogonal-multigrid-multilevel-monte-carlo}. Numerical results are then reported in Section~\ref{sec:experiments}.

\subsection{Physical Applications} \label{subsec:physical_applications}

The first application that we will address in the numerical experiments is the computation of the time-averaged pseudoscalar connected two-point function

\begin{equation}
\label{eq:connected-G}
G(t) = \frac{1}{T} \sum_{t'=1}^{T} \mathrm{tr}\left( D^{-1}(t+t',t')\,\Gamma_{5}\,D^{-1}(t',t+t')\,\Gamma_{5}\right) \,.
\end{equation}

Here, $D^{-1}(t_i,t_j)$ denotes the block of the propagator that maps the time slice $t_j$ to the time slice $t_i$. 
The trace in \cref{eq:connected-G} therefore runs only over color, spin, and spatial indices at fixed times, not over the full four-dimensional lattice.  This operator is the standard connected pseudoscalar two-point function at zero momentum, which is widely used in lattice QCD to extract the pion mass and decay constant; see, for example, \cite{mesons}.

As a second application, we will address in \cref{sec:experiments} is the trace of the inverse of the full Dirac operator, as a function of a mass shift $m$. The trace $\tr[(D+m)^{-1}] = \sum_x (D+m)^{-1}(x,x)$, where $x$ denotes the four-dimensional lattice sites $(\vec{x},x_0)$, has to be computed to obtain $\frac{d}{dm} \langle O \rangle^{\rm gauge}$, which enters the first-order Taylor expansion:
\begin{equation}
    \langle O \rangle^{\rm gauge}_{m'} = \langle O \rangle^{\rm gauge} + (m'-m) \frac{d}{dm} \langle O \rangle^{\rm gauge} ,
\end{equation}
used to correct small mistunings of the bare quark mass $m'$ after gauge configurations have been generated at mass $m$ \cite{Bruno:2016plf,Hollwieser:2020qri,Strassberger:2021tsu}.

A third application is the trace of the inverse of the discretized Dirac operator on a time slice multiplied by an operator $\Gamma$, $\tr(\Gamma D^{-1}(t,t))$. This quantity enters the calculation of disconnected fermion loop contributions\footnote{\textit{Disconnected} refers to fermionic loops, where fermion lines start at a spacetime point $x$ and return to the same point $x$, forming a closed loop \cite{Gattringer2010}. For flavor singlet mesons the expression in \cref{eq:connected-G} receives an additional contribution proportional to the product of the traces of two such loops at times $t$ and $t'$ respectively.}. The operator $\Gamma$ fixes the symmetry channel. For example, choosing $\Gamma=\Gamma_{5}$ gives the pseudoscalar channel, which appears in the two-point function of flavor-singlet pseudoscalar mesons.

\section{Variance Reduction Techniques}
\label{sec:variance-reduction-techniques}

We start by reviewing some of the widely used techniques to reduce the variance of the Hutchinson estimator \eqref{eq:Hutchinson}, namely (exact or inexact) deflation of singular vectors (\cref{sec:Deflation}), as well as various multilevel Monte Carlo approaches (\cref{sec:multigrid-multilevel-monte-carlo}). We then introduce in \cref{sec:orthogonal-multigrid-multilevel-monte-carlo} a novel modification of multigrid multilevel Monte Carlo, the guiding idea of which is to replace oblique projections by {\em orthogonal} ones.

\nopagebreak
\subsection{Deflation}
\label{sec:Deflation}

Given the relation between the Frobenius norm and the singular values $\sigma_i$ of a matrix $A$ (see, e.g. \cite{Trefethen1997}):

\begin{equation}
\label{eq:Frobenius-Singular}
   ||A||_F^2 = \sum_{i=1}^n \sigma_i^2,
\end{equation}
several approaches to reduce the variance of the Hutchinson estimator have focused on \textit{deflating} the operator to remove contributions of the $k$ largest singular values $\sigma_1 \geq \sigma_2 \geq \ldots \geq \sigma_k$ of $A$ to the variance \cite{Gambhir2017, Hutch++}. The underlying heuristic is that reducing $\| A\|_F$ also reduces $\|\text{offdiag(A)}\|_F$ and similar quantities that enter the variance for the different probability distributions; see \cref{eq:variance-Frobenius-Relation}, e.g.

 Following \cite{Gambhir2017}, let $A=U\Sigma V^{\dag}$ be the singular value decomposition (SVD) of $A$. 
 The columns of the unitary matrices $U$ and $V$ are the left and right singular vectors, respectively, and 
 $\Sigma = \diag(\sigma_1,\ldots,\sigma_n)$
 is the diagonal matrix of singular values. Let $U = [U_k, 
 U_{\neg k}]$, $V = [V_k, V_{\neg k}]$ where $U_k, V_k$ hold the first $k$ columns of $U$ and $V$, respectively, 
 representing the $k$ left and right singular vectors of 
 $A$ belonging to the $k$ largest singular values. The orthogonal projector $\Pi_k = U_kU_k^{\dag}$ deflates 
 $A$ in the sense that the singular value decomposition of $(I-\Pi_k)A$ is $U_{\neg k}\Sigma_{\neg k}V_{\neg k}^\dag$ with $\Sigma_{\neg k} = \diag(\sigma_{k+1},\ldots,\sigma_n)$ and, similarly, $\Pi_k A = U_k \Sigma_k V_k^\dag$ with $\Sigma_k = \diag(\sigma_1,\ldots,\sigma_k)$.
 
 One then splits the $\tr(A )$ in the two terms
\begin{equation}
\label{eq:deflation-general}
        \tr(A) = \tr \left( (I - \Pi) A  \right) + \tr (\Pi A ). 
\end{equation}

The first term contains the singular modes corresponding to the remaining smaller $n-k$ singular values of $A$, reducing the Frobenius norm to 
\begin{equation}
\label{eq:Frobenius-reduction-singular}
    || (I - \Pi) A  ||_F^2 = || A ||_F^2 - \sum_{i=1}^k \sigma_i ^{2}.
\end{equation}
Due to \cref{eq:variance-Frobenius-Relation} we thus expect a reduction in the variance when estimating  $\tr \left( (I - \Pi) A \right)$ with Hutchinson's method. Using the cyclic property of the trace\footnote{$\tr(XY) = \tr(YX)$ for any matrices $X, Y$ for which the products are defined}, the second term in \ref{eq:deflation-general} can be expressed as $\tr (\Pi A) =  \tr ( U_k^{\dag} A U_k )$. Since $AV_k = \Sigma_k U_k,$ we obtain this trace directly and thus non-stochastically, using the precomputed singular modes and values, as  

\begin{equation}
   \tr (\Pi A) = \tr(V_k^\dag \Sigma_k U_k)= \sum_{i=1}^k {\sigma_i} v_i^{\dag} u_i.
\end{equation}

We note that the reduction of the Frobenius norm described in eq. (\ref{eq:Frobenius-reduction-singular}) represents the largest reduction achievable among all rank-$k$ projections (see, e.g., \cite{Trefethen1997}). 
It is worth mentioning that the multilevel deflation scheme we introduce later uses a projector of much larger rank (proportional to the matrix dimension).

Let us also note that in the situation where $A = B^{-1}$, the SVD $B = U\Sigma V^\dag$ of $B$ gives the SVD $A = V\Sigma^{-1}U^\dag$, which implies that the projector $\Pi_k = V_k V_k^\dag$ is constructed from the $k$ {\em right} singular vectors of $B$ belonging to the $k$ smallest singular values of $B$.

A modification of the above {\em exact} deflation approach is {\em inexact} deflation where $U_k$ are allowed to only hold, possibly quite inaccurate, approximations to the left singular vectors of $A$. In the case of $A$ being a matrix inverse, $A = B^{-1}$, the term $\tr ( V_k^{\dag}A V_k)$ can still be computed non-stochastically by solving the $k$ linear systems with matrix $B$ and the columns of $V_k$. 

Instead of employing an orthogonal projector $\Pi = V_kV_k^{\dag}$, one also has the option of constructing an oblique projector, i.e., $\Pi^2 = \Pi$, but $\Pi^{\dag}\neq \Pi$. This choice is referred to as oblique deflation in the  
literature~\cite{ObliqueDef}.

\subsection*{Deflation for the Dirac operator}

The Dirac operator $D$ is non-Hermitian but satisfies the $\Gamma_5$-Hermiticity property, i.e.\ $Q^\dag = Q$ for $Q = \Gamma_{5}D$ with $\Gamma_5$ being a simple diagonal matrix which flips the sign of half of the spin components. The eigendecomposition $Q =V\Lambda V^{\dag}$ of $Q$ gives the SVD of $D$ as $D = U \Sigma V^{\dag}$, with $U = \Gamma_5 V\,\mathrm{sign}(\Lambda)$ and $\Sigma = \mathrm{abs}(\Lambda)$.

Thus, for exact or inexact deflation, one can compute and use $k$ exact or approximate eigenvectors $v_i$ corresponding to the small eigenvalues of $Q$. They can be obtained, for example, via the block power or block Rayleigh-Ritz method applied to $Q^{-1}$. The resulting deflated Hutchinson trace estimator for $\tr(D\mo)$ with sample size $N$ is then
\begin{equation}
\widehat{\tr}(D^{-1}) = \frac{1}{N}\sum_{n=1}^{N}(x^n)^\dag\left(I-V_kV_k^\dag\right) D^{-1} x^n
+\tr(V_k^\dag D^{-1} V_k),\label{eq:deflated-estimator}
\end{equation}
where $V_k = [v_1| \cdots |v_k]$.

\subsection{Multilevel Monte Carlo}
\label{sec:multigrid-multilevel-monte-carlo}

The idea underlying multilevel Monte Carlo is to additively decompose a stochastic variable $X$ into $L$ stochastic variables $X_l$ as
\[
X = \sum_{l=1}^L X_l
\]
in such a way that $\V(X_l)$ is small when the cost $C_l$ to compute a sample for $X_l$ is high. One can then minimize the total cost for achieving a target variance $\epsilon^2$ for the sum of the sample means
\[
\sum_{l=1}^L \frac{1}{N_l} \sum_{k=1}^{N_k} X_l^{(k)},
\]
by taking 
\begin{equation}\label{eq:MLMC-Nl}
N_l = \frac{1}{\epsilon^2} \sqrt{\frac{V_l}{C_l}} \sum_{i=1}^L \sqrt{V_i C_i}.
\end{equation}
We refer to \cite{Giles2015} for a derivation of this result and for an in-depth discussion of the general multilevel Monte Carlo principle.

Various multilevel Monte Carlo methods have been considered for the computation of the trace of inverse operators arising in lattice QCD. Generically, they all start from a telescopic sum decomposition of the inverse operator $A$
\[
A = \sum_{l=1}^{L-1} (A_{l}-A_{l+1}) + A_L,
\]
with $A_1 = A$. Then
\begin{equation*}
\tr(A) = \sum_{l=1}^{L-1} \tr(A_{l}-A_{l+1}) + \tr(A_L),
\end{equation*}
and we can use (standard or deflated) Hutchinson on each of the terms on the right. The matrices $A_l$ should be chosen such that the cost to evaluate $(A_{l}-A_{l+1})x$ or $A_Lx$ is small when the variance is high and that the variance is small for those where the cost is high.

In the {\em frequency splitting} approach of \cite{Giusti2019} ($A = D^{-1}$) one takes $A_l = (D+\sigma_l I)^{-1}$, with carefully chosen shifts $\sigma_l \in \mathbb{C}$; see, e.g., \cite{Whyte202410892}. 
The {\em truncated solver method} \cite{Bali2010} (again $A=D^{-1}$) is based on a convergent splitting $D=M-N$ of $D$ and takes $A_l = \sum_{j=1}^{k_l}(M^{-1}N)^jM^{-1}$ with $k_2 \geq k_3 \geq \cdots \geq k_L$, which is an increasingly better approximation to $D^{-1}$ the larger $k_l$, and where $A_lx$ amounts to performing $k_l$ steps of the stationary iteration $Mx^{(j)} = Nx^{(j-1)}+x$, $j=1,\ldots,k_l$, $x^{(0)} = 0$. {\em Polynomial} \cite{Baral2019} and {\em multipolynomial} \cite{Lashcomb2024} multilevel Monte Carlo approaches take $A_l$ as polynomials of degree $d_l$ in $D$, approximating $D^{-1}$. The approximation is more precise when $d_l$ is large and the degrees $d_l$ satisfy $d_2 \geq d_3 \cdots \geq d_L$. 

Finally, in {\em multigrid} multilevel Monte Carlo for $\tr(D^{-1})$ \cite{FrommerMultilevel,Hallman-Troester2022}, see also \cite{Bali2015,MultigridDeflation,Gruber-MG-LMA}, one uses the operator hierarchy available from a multigrid solver for $D$. 
We denote by $D_l$ the operator at level $l$, $D_l \in \mathbb{C}^{n_l \times n_l}, D_1 = D$ and by $R_l$ and 
$P_l$ the restriction and prolongation grid transfer operators between levels $l$ and $l+1$. Moreover, we use $\hat{P}_l, \hat{R}_l$ for the accumulated grid transfer operators $\hat{P}_l = P_1\cdots P_l$, and $\hat{R}_l = R_l\cdots R_1$ between levels 1 and $l+1$. In multigrid multilevel Monte Carlo, one now takes
\begin{equation} \label{eq:Al_def:eq}
A_l = \hat{P}_{l-1}D_{l}^{-1}\hat{R}_{l-1}.
\end{equation}

All multigrid methods for the Dirac operator, like DD$\alpha$AMG \cite{DDalphaAMG-paper}, the method available in QUDA \cite{CLARK-Quda-2010} based on \cite{Babich2012}, or the OpenQCD solver \cite{openQCD} are adaptive and aggregation based. In particular, they all take $R_l = P_l^\dagger$ with unitary $P_l$, i.e.\ $P_l^\dagger P_l = I$. As a consequence, due to the cyclic property of the trace, we have $\tr(A_l-A_{l+1}) = \tr(D_l^{-1}-P_lD_{l+1}^{-1}R_l)$, which effectively reduces the dimensions and thus the cost when applying Hutchinson's estimator to this contribution in the telescopic sum.    

The coarse grid operators are obtained via a Galerkin construction, i.e.\ $D_{l+1} = R_lD_lP_l$. Thus, the operator $\Pi_l = P_lD_{l+1}^{-1}R_lD_l$ is a projection, and we have
\begin{equation}
\label{eq:oblique-deflation}
 M_l := (I-\Pi_l)D_l^{-1} = D_l^{-1} - P_lD_{l+1}^{-1}R_l,
\end{equation}
showing that we can interpret each term $D_l^{-1} - P_lD_{l+1}^{-1}R_l$ arising in multigrid multilevel Monte-Carlo as resulting from a deflation using the oblique projector $\Pi_l$. The rationale behind multigrid multilevel Monte Carlo is that due to the aggregation based construction of the transfer operators and a property of the low eigenmodes called ``local coherence'' \cite{Luscher:Local-Coherence}, many of the prolongated low modes of $D_{l+1}$ are good approximations to those of $D_l$. Consequently, many of the large modes of $D_l^{-1}$ are approximately removed when subtracting the coarse-grid correction, effectively deflating the near kernel. In practice, the transfer operators $P_l$ and $R_l$ are constructed during a \emph{setup phase} from $N_{\mathrm{tv}}$ \emph{test vectors} $v_1,\ldots,v_{N_{\mathrm{tv}}} \in \mathbb{C}^{n_l}$ that approximate eigenvectors associated with its smallest eigenvalues of $D_l$. Summarizing, the multigrid Multilevel Monte Carlo decomposition for $\tr(D^{-1}$ is
\begin{equation} \label{eq:oblique_multigrid-multilevel-monte-carlo}
    \tr(D^{-1} = \sum_{l=1}^{L-1} \tr(M_l) + \tr(D_L^{-1}).
\end{equation}

An application of multigrid multilevel Monte Carlo was recently used in \cite{Gruber-MG-LMA} where the multilevel deflation in \cref{eq:oblique-deflation} was applied within the Wick contraction of hadronic two-point correlators. This achieved speed-ups of up to 30 compared with the plain-Hutchinson method. The construction of the multigrid hierarchy in \cite{Gruber-MG-LMA} used many ($N_{tv}=50$), \emph{exact} eigenvectors of the Hermitian operator $\Gamma_5 D$ as test vectors. Our results in section~\ref{sec:connected_case} show that we achieve variance reductions of the same order with a smaller number of test vectors, which, in addition, may be only moderately accurate eigenvectors, and thus with a reduced setup cost.

\section{Multigrid Multilevel Monte Carlo Relying on Orthogonal Projections}
\label{sec:orthogonal-multigrid-multilevel-monte-carlo}

If $\Pi$ is an orthogonal projector, then $\|\Pi A \|_F \leq \| A \|_F$, an inequality that need not be satisfied when $\Pi$ is oblique. 

When using projections in a variance reduction technique for the Hutchinson estimator, given the dependence of the variance on the Frobenius norm, the question arises whether we can devise a modification of multigrid multilevel Monte Carlo which uses orthogonal projections, and whether this then results in a larger reduction of the variance.  This is what we address now. 

We take the (unitary) prolongation operators $P_l$ of the multigrid construction to build the orthogonal projector $\Pi_l = P_l P_l^{\dag}$. This allows one to deflate the operator $D_l\mo$ as 

 \begin{equation}
    \label{eq:orthogonal-decomposition}
    D_l\mo = (I-P_lP_l^{\dag})D_l\mo + (P_lP_l^{\dag}D_l\mo), \quad l = 1, \ldots L.
\end{equation}

Computing the trace for the second term can be transferred to the next coarser level, by using the cyclic property of the trace as

\begin{equation}
    \tr (P_lP_l^{\dag} D_l\mo) = \tr (P_l^{\dag} D_l\mo P_l - D_{l+1}\mo) + \tr(D_{l+1}\mo),
\end{equation}
\\
so that $\tr(D_l\mo)$ is split into three terms 

\begin{equation} 
    \label{eq:orthogonal_multilevel_notrick}
    \tr(D_l\mo) = \tr((I-P_lP_l^{\dag})D_l\mo) +\tr(P_l^{\dag}D_l\mo P_l-D_{l+1}\mo)+ \tr(D_{l+1}\mo),
\end{equation}
and we can recurse on the third term. 

Regarding the first term $\tr((I-P_lP_l^{\dag})D_l\mo)$ in \cref{eq:orthogonal_multilevel_notrick}, we can use the fact that $(I-P_lP_l^{\dag})^2 = I-P_lP_l^\dag$, and the cyclic property of the trace, to obtain the two-sided projected variant 
\[
\tr((I-P_lP_l^{\dag})D_l\mo = \tr((I-P_lP_l^{\dag})D_l\mo(I-P_lP_l^{\dag})).
\]
The additional factor of $I-P_lP_l^{\dag}$ on the right further reduces (more precisely: is guaranteed to not increase) the Frobenius norm and is thus likely to result in a lower variance of the Hutchinson's estimator. Computationally, the cost for one stochastic estimate remains the same: We compute $y^{(k)} = (I-P_lP_l^\dagger) x^{(k)}$ once, do one solve for $Dz^{(k)} = y^{(k)}$, and then take $(y^{(k)})^\dag z^{(k)}$, whereas in the one-sided variant we take $(z^{(k)})^\dag x^{(k)}$. 

In the second term in \cref{eq:orthogonal_multilevel_notrick}, the matrix  $P_l^{\dag}D_l\mo P_l$ has full rank $n_{l+1}$, which bears the potential that $P_l^{\dag}D_l\mo P_l-D_{l+1}\mo$ be comparatively small so that the Hutchinson estimator for its trace will have a small variance. We will see in section~\ref{sec:experiments} that this is indeed the case.

Using the notation
\begin{equation} \label{OandF_def:eq}
\left.
\begin{array}{rcl}  
O_l &:=& (I-P_lP_l^{\dag})D_l\mo(I-P_lP_l^\dag ), \\ 
F_l &:=& P_l^{\dag}D_l\mo P_l-D_{l+1}\mo
\end{array}
\quad \right\}
\end{equation}
for the two-side projected operator and the difference of the full rank operators, respectively, the new multilevel decomposition for the trace is 
\begin{equation}
\label{eq:smultigrid-multilevel-monte-carlo}
     {\tr}(D^{-1})  = \sum_{l=1}^{L-1} {\tr}(O_l) + \sum_{l=1}^{L-1} {\tr}(F_l) + {\tr}(D_L\mo).
\end{equation}

We obtain an unbiased estimator by using Hutchinson's method on each of its terms, possibly using a direct, non-stochastic computation for ${\tr}(D_L\mo)$.

\section{Numerical Experiments} \label{sec:experiments}

We present three complementary studies to quantify the variance reduction delivered by multigrid multilevel Monte Carlo, comparing with the plain Hutchinson estimator and with inexact deflation. We begin with the connected pseudoscalar correlator $G(t)$ 
then turn to the trace $\tr(D^{-1})$ of the full Dirac operator 
(\cref{sec:4D_case}), and finally look at the case of $\tr(\Gamma_5D\mo(t,t)$. 

The variances reported are estimated using the unbiased estimator given by the sample mean of a complex random variable $X$ with estimated mean $\hat{\mu}$, i.e.\ we take the estimator $\widehat{\V}[X] = \frac{1}{N-1} \sum_{i=1}^{N} |X_i - \hat{\mu}|^2$. The methods require solves with level operators $D_l$ which we do using the DD$\alpha$AMG multigrid solver starting at the respective level.  At all levels, including the coarsest, we solve the respective linear systems with the accuracy requirement that the relative residual be reduced by $10^{-12}$.
The test vectors used to construct the transfer operators in DD$\alpha$AMG are computed as very rough approximations (with residuals of order $10^{-1}$) to small eigenvalue modes of $D$. These low-accuracy vectors are sufficient to capture the near-kernel space required for the multigrid solver.

\subsection{Connected pseudoscalar two-point function}
\label{sec:connected_case}

We assess the variance reduction in  multigrid based multilevel decompositions of the connected operator $G(t)$ defined in \cref{eq:connected-G}, using a configuration from the IV ensemble generated by the Regensburg QCD (RQCD) collaboration \cite{Regensburgcollab}; see \cref{tab:configurations}.

Similarly to what was done in \cite{Gruber-MG-LMA} for the oblique projection case, 
we define deflated operators, using the notation $D_l^{-1}(t+t',t')$ for the $(t+t',t')$ time-slice block of $D_l^{-1}$ (rows at $t+t'$, columns at $t'$) as 
\begin{eqnarray*}
\label{eqn:B_l}
B_l(t+t',t') &=& \hat{P}_l(D_l\mo(t+t',t') -P_lD_{l+1}\mo(t+t',t') P_l^\dag)\hat{P}_l^\dag, \enspace l=1,\ldots,L-1,    \\
B_L(t+t',t') &=& \hat{P}_LD_{L}\mo(t+t',t')\hat{P}_L^\dag.
\end{eqnarray*}
With this,  the correlator in \cref{eq:connected-G} can be written as
\begin{equation}
\label{eq:G-traditional}
    G(t) = \sum_{i,j=1}^{L}G_{i,j}(t) ,
\end{equation}
where
\begin{equation}
\label{eq:G-ij}
    G_{i,j}(t) = \frac{1}{T} \sum_{t'=1}^{T}  \tr \big(B_i(t+t',t')\Gamma_5 B_j(t',t+t')\Gamma_5 \big ).
\end{equation}

Note that the presence of the $\Gamma_5$ factors prevents the use of the cyclic property of the trace, which was possible for the full Dirac propagator $D\mo$. As a result, the sequence of the accumulated transfer operators $\hat{P}_l$ and $\hat{P}_l^H$ has to be kept.

For an orthogonal projection approach orthogonal we start from  \cref{eq:orthogonal-decomposition} to obtain
\begin{equation*}
    D_l\mo = \underbrace{(I - P_lP_l^\dag) D_l\mo }_{J_l} + \underbrace{P_l P_l^\dag D_l\mo  - P_lD_{l+1}\mo P_l^\dag}_{K_l}  + P_lD_{l+1}\mo P_l^\dag P_l.
\end{equation*}
and define the deflated operators
\begin{equation} \label{eq:full_ops_orth}
\left. \begin{array}{rcll}
B_{l} &:=& \hat {P}_l\, J_l\, \hat{P}_l^\dag
, & l = 1,\ldots,L-1, \\
B_{L-1+l} &:=& \hat P_{\,l+1}\, K_l\, \hat{P}_{\,l+1}^\dag, & l = 1,\ldots,L-1, \\
B_{2L-1} &:=& \hat P_L\, D_L^{-1}\, \hat P_L^\dag, &
\end{array} 
\enspace \right\}.
\end{equation}
For $r,s \in \{1,\ldots,2L-1\}$ we set 
\[
G_{r,s}(t)
:= \frac{1}{T}\sum_{t'=1}^{T} \tr \big( B_{r}(t+t',t')\,\Gamma_5\, B_{s}(t',t+t')\,\Gamma_5\big ),
\]
obtaining a decomposition of $G(t)$  in \cref{eq:connected-G} as
\begin{equation}
    \label{eq:G-split}
     G(t) = \sum_{r,s = 1}^{2L-1}  G_{r,s}(t).
\end{equation}
Note that the number of summands has increased from $L^2$ to $(2L-1)^2$ compared to the  oblique case.

\begin{table}[h]
\centering
\caption{Configurations analyzed in this section. For each configuration we list the lattice size and the sizes of the coarser lattices (levels 2--4) used in the multigrid hierarchy.}
\label{tab:configurations}
\begin{tabular}{lcccc|cc}
\toprule
Ensemble & $T \times L^3$ & Level 2 & Level 3 & Level 4 & $m_\pi$ [MeV] & $a$ [fm] \\
\midrule
E250 & $192 \times 96^3$ & $48 \times 24^3$ & $24 \times 12^3$ & $12 \times 6^3$ & $131$ & 0.063 \\
J501 & $192 \times 64^3$ & $48 \times 16^3$ & $24 \times 8^3$  & $12 \times 4^3$ & $336$ & 0.039 \\
D450 & $128 \times 64^3$ & $32 \times 16^3$ & $16 \times 8^3$  & --              & $216$ & $0.075$ \\
IV   & $64 \times 32^3$  & $16 \times 8^3$  & $8 \times 4^3$   & --              & $295$ & 0.071 \\
\bottomrule
\end{tabular}
\end{table}

We estimate with a fixed sample size $N = 128$ the variance of the plain Hutchinson estimator of $G(t)$ by stochastically computing each of the terms of $G(t)$ in \cref{eq:connected-G}. We also do this for the $G_{i,j}$ terms of the oblique and orthogonal multilevel Monte Carlo decompositions from \cref{eq:G-traditional} and \cref{eq:G-split}. Note that in the oblique projection case, each operator $B_l$ involves solves at levels $l$ and $l+1$, except for $B_L$, which uses only the coarsest-level solve. As a result, the indices of $G_{i,j}$ indicate the levels involved in computing the term. For example, $G_{1,1}$ requires solves at levels 1 and 2, while $G_{2,3}$ involves solves at level 2 and at the coarsest level. A similar interpretation applies in the orthogonal case, where the levels involved in each $G_{r,s}$ term depend on the corresponding projected operators.

\begin{figure}[h]
\centering
\includegraphics[scale=0.35]{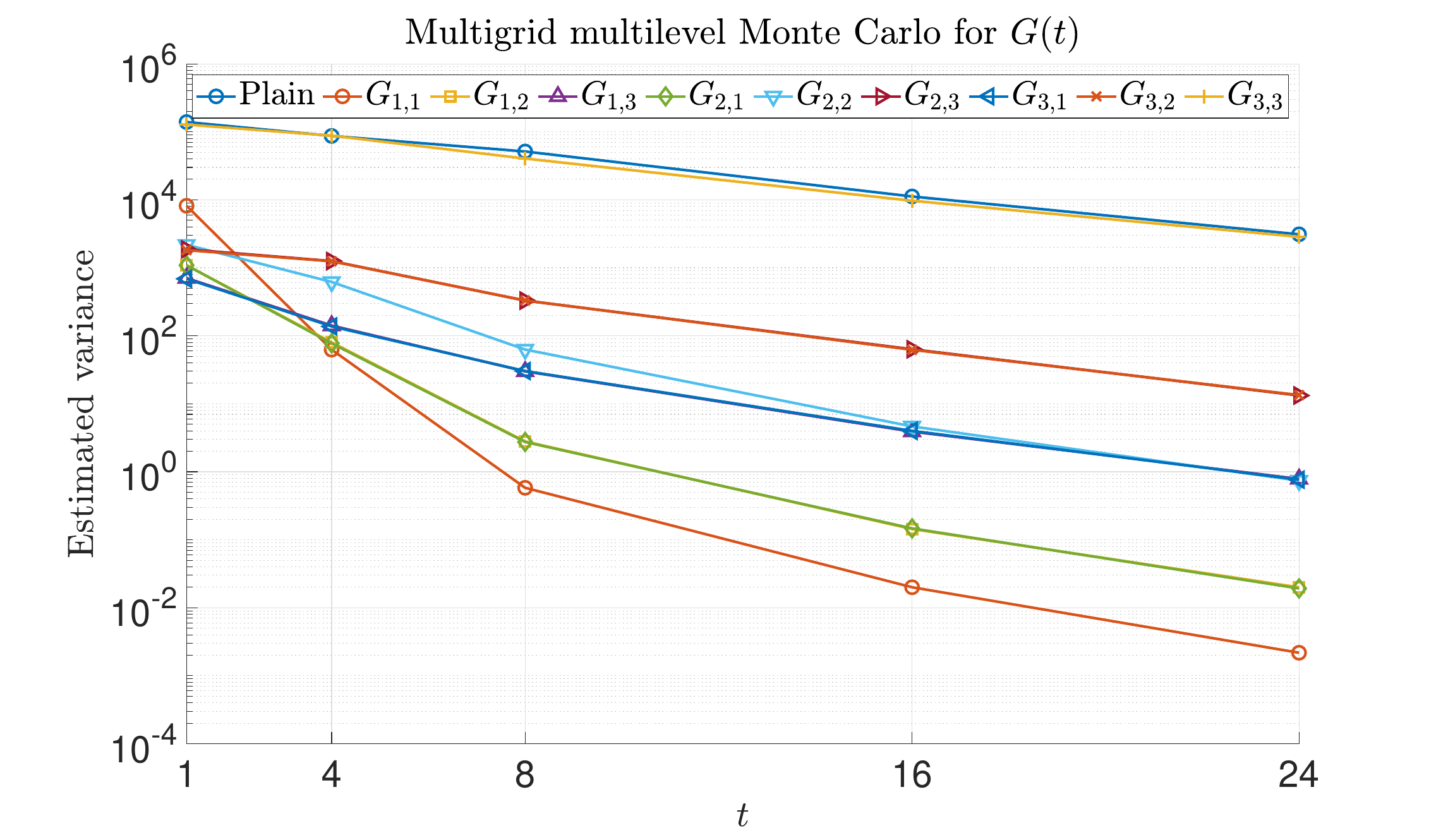}
\caption{\label{fig:connected-traditional} Variance reduction for the connected two-point function, IV configuration. Comparison between the plain Hutchinson estimator and the multigrid Multilevel Monte Carlo estimator based on oblique projections.}
\end{figure}

\Cref{fig:connected-traditional} presents the variance of the plain Hutchinson estimator of $G(t)$ and compares it to the $G_{i,j}$ terms of the multilevel decomposition based on oblique projections from \cref{eq:G-traditional}. 
It shows that $G_{1,1}$, the most expensive term, achieves a dramatic variance reduction relative to the plain Hutchinson estimator (top blue curve): at $t=1$ the reduction is already on the order of $10^3$, and grows systematically as $t$ increases and, by $t=24$, it reaches roughly $10^5$. At the opposite end, the cheapest term, $G_{3,3}$, carries the largest variance among the $G_{i,j}$ components across all $t$. 
This distribution of the variances is favorable for multilevel Monte Carlo as the variance is concentrated in the inexpensive coarsest term, which can be sampled with a low cost. Overall, the decomposition is such that the variance is reduced as the cost grows and vice versa.

\begin{figure}[h]
\centering
\includegraphics[width=1.0\linewidth]{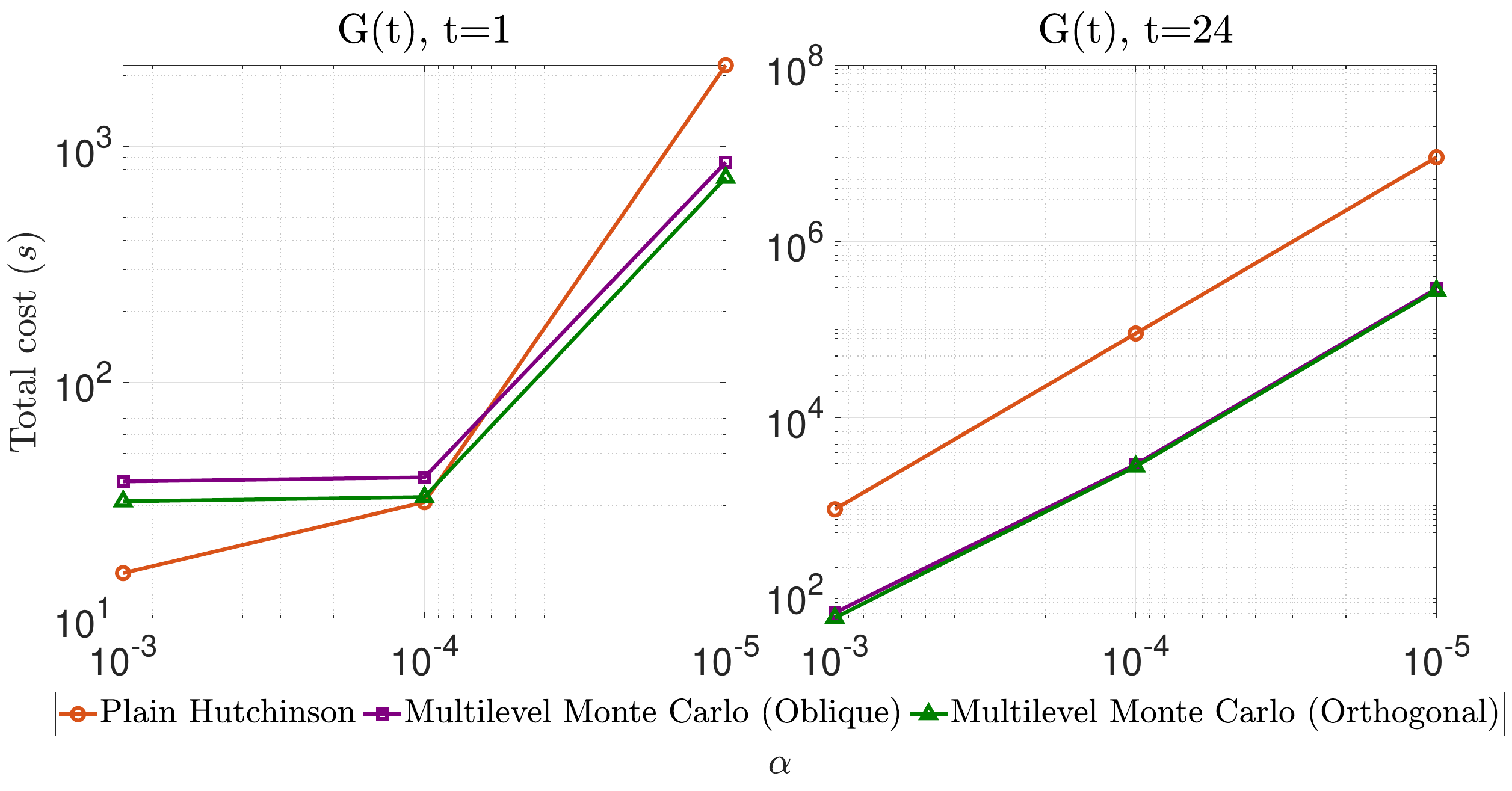}
\caption{\label{fig:connected-cost}
Total cost  for the connected two-point function on the IV configuration. We compare between the plain Hutchinson estimator and the two multigrid multilevel Monte Carlo variants using oblique and orthogonal projectors.}
\end{figure}

\Cref{fig:connected-cost} summarizes the total cost needed to attain a relative accuracy $\alpha$ when estimating $G(t)$ at two representative separations, $t=1$ and $t=24$, comparing the plain Hutchinson estimator with the oblique formulation \cref{eq:G-traditional} and the orthogonal formulation \cref{eq:G-split}. At $t=1$ (left), the cost of the plain Hutchinson is the lowest for the relaxed target relative accuracy of $10^{-3}$. The reason is that in this case only one sample is enough to achieve the desired accuracy. When higher accuracies are needed, the multilevel Monte Carlo decompositions perform better, as the smaller variances they have entail a reduced sample size. Furthermore, since $G(t)$ decreases with $t$, and the variance of the plain Hutchinson estimator remains roughly constant, the variance reduction achieved by the multilevel methods becomes increasingly significant at larger separations. At $t=24$, this results in a cost reduction of up to a factor of 30 when using our multilevel Monte Carlo methods.

The decomposition from \cref{eq:G-split}, which is based on orthogonal projectors, yields a consistently cheaper estimation of $G(t)$ compared to the oblique projector–based decomposition from \cref{eq:G-traditional}. This is notable, as the orthogonal approach involves 25 terms in the three-level method used, whereas the oblique method includes only 9. Nevertheless, the orthogonal variant achieves an overall cost reduction of 5–15\%, due to the significantly stronger variance reduction in each $G_{s_i,d_j}$ term, which compensates for the larger number of contributions.

\subsection{Inverse Dirac operator} \label{sec:4D_case}

\begin{figure}[h]
\hspace{-0.5cm} \includegraphics[scale=0.29]{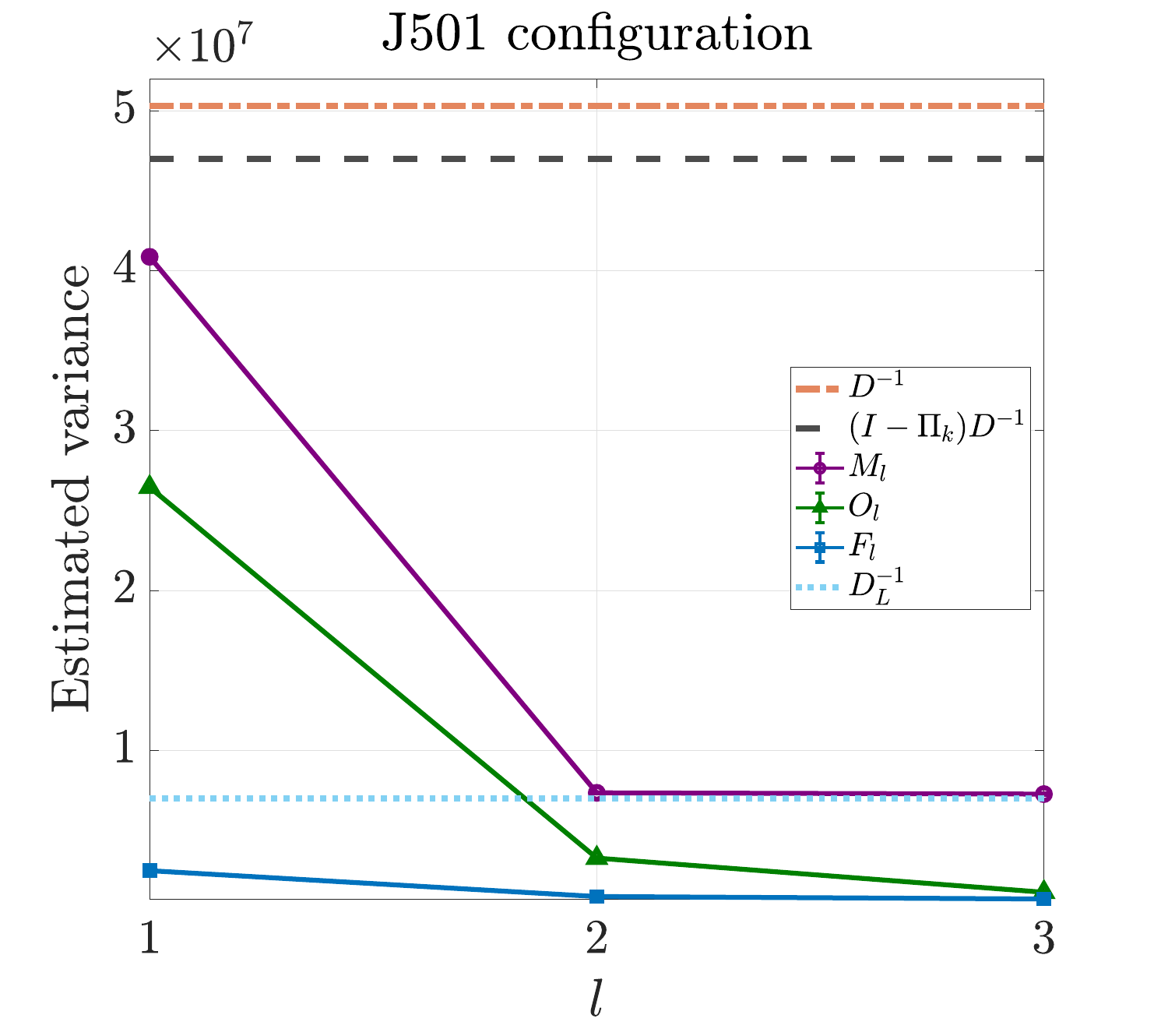}
\includegraphics[scale=0.29]{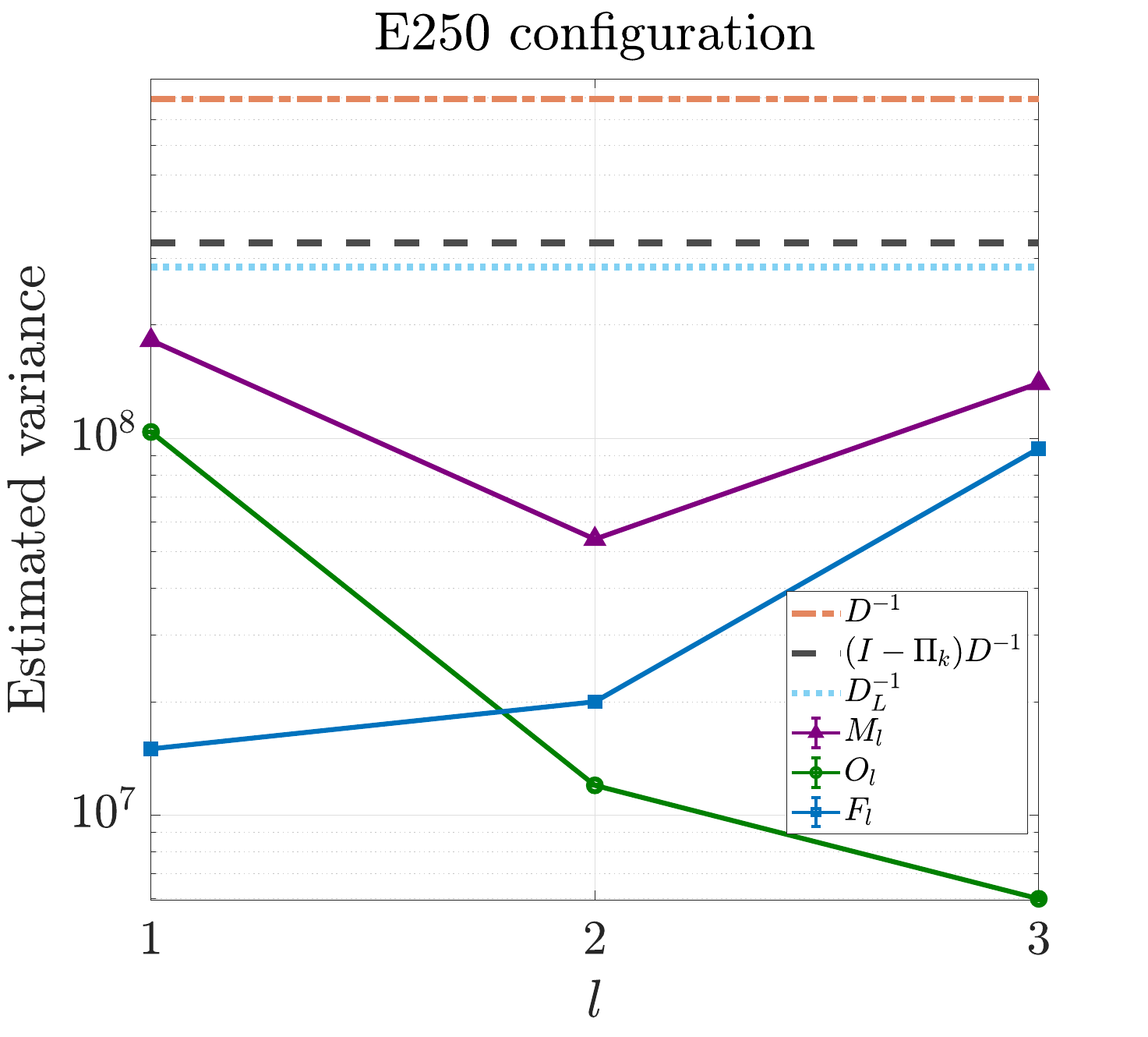}
\caption{\label{fig:E250-J501-variances}
Variance reduction for $\tr(D^{-1})$ for the J501 and E250 configurations. Variances are estimated with a fixed sample size of $N=500$. Shown are the contributions from inexact deflation $(I - \Pi_k) D^{-1}$ ($k=128$), the multilevel oblique decomposition ($M_l$), the full-rank terms ($F_l$), the orthogonal decomposition ($O_l$), and the coarsest-level term $D_L^{-1}$, evaluated across levels $l$.}
\end{figure}

We now turn to evaluating the variance reduction in estimating  $\tr(D^{-1})$, the trace of the inverse of the full Dirac operator, using four levels when building the multigrid hierarchy. For this case, multigrid Multilevel Monte Carlo can directly be built on the decomposition \cref{eq:oblique_multigrid-multilevel-monte-carlo} involving the matrices $M_l$ from eq.~\eqref{eq:oblique-deflation} (oblique projections) and \eqref{eq:smultigrid-multilevel-monte-carlo} with the matrices $O_l,F_l$ from eq.~\eqref{OandF_def:eq} (orthogonal projections). Results are shown for configurations of two ensembles from the CLS Collaboration \cite{Strassberger:2021tsu}, J501 and E250, whose lattice sizes and the corresponding multigrid setups are listed in \cref{tab:configurations}. All variance measurements are evaluated with a fixed sample size of $N=500$ and variance contributions are computed for all levels $l = 1, 2, 3$ and $L=4$. \Cref{fig:E250-J501-variances} compares the variance contributions of the plain Hutchinson method, inexact deflation, and both multigrid multilevel Monte Carlo variants: 
the one using oblique projectors and the matrices $M_l$ from eq.\ \eqref{eq:oblique-deflation} and the one using orthogonal projectors and the matrices $O_l$, $F_l$ from eq.\ \eqref{OandF_def:eq}.

In the inexact deflation method of \cref{eq:deflated-estimator}, the deflation basis $V_k$ is constructed using five iterations of an inverse block power method applied to $\Gamma_5D$, with each inner inversion converging to a relative residual of $10^{-6}$. The second term in the estimator is then computed with the required $k$ solves and $k$ vector products.

On the J501 configuration, the orthogonal decomposition achieves the highest reduction at the finest (and most expensive) level: the variance for $O_1$ lies below that of the other methods, and the variance for $F_1$ (which has full rank) is about one order of magnitude smaller than for $O_1$. The variances for both  $O_l$ and $F_l$ decrease toward the coarse levels, while the variance for coarsest level operator remains small. Overall, the oblique variant exhibits a better variance reduction than inexact deflation. Moreoever, the variances for $M_l$ remain above those for $O_l$ at all levels.

For the more ill-conditioned E250 ensemble, this behavior persists. The worse conditioning is evident from the increased variance of both the plain Hutchinson and the inexactly deflated estimator. Inexact deflation now achieves a visibly larger reduction than on J501. The orthogonal variant of multigrid Multilevel Monte Carlo achieves the highest variance reduction at the finest level. Moreover, the variance of the $O_l$ terms decreases toward coarser levels while, at the same time, the variance of the $F_l$ terms grows. This higher reduction is also reflected in the $D_L^{-1}$ term, which is the cheapest to compute and, in this case, also carries the highest variance in the multilevel decomposition.

While these results for $\tr(D\mo)$ are illustrative, the overall variance reductions achieved in this trace estimation are modest compared to those obtained for the connected correlator $G(t)$ from section~\ref{sec:connected_case}. In particular, the gains are not sufficient to make the multilevel estimators more cost-effective than the plain Hutchinson method in this case. However, the results do highlight a trend: the more ill-conditioned the configuration, the more effective the multilevel variance reduction becomes. This suggests that the method may be particularly beneficial for twisted mass ensembles, where conditioning is typically worse. For practical applications in such settings or where larger gains are required, we think that alternative strategies, such as combination with probing, may be beneficial.

\subsection{Disconnected fermion loop contributions}

Our last example considers $\tr(\Gamma_5 D^{-1}(t,t))$, which occurs when computing certain disconnected fermion loop contributions. To obtain a multilevel hierarchy we use the sum 
\begin{equation}
\label{eq:trace_equality_disconnected}
\tr(\Gamma_5(t) D^{-1}(t,t)) = \sum_{l=1}^L \tr \left( \Gamma_5 B_l(t,t) \right)
\end{equation}
with $B_l$ from \cref{eqn:B_l} in the oblique projection case and we could proceed similarly with the matrices defined in \eqref{eq:full_ops_orth} for the orthogonal projection case.

\begin{figure}[h]
\hspace{-0.5cm}  \includegraphics[scale=0.26]{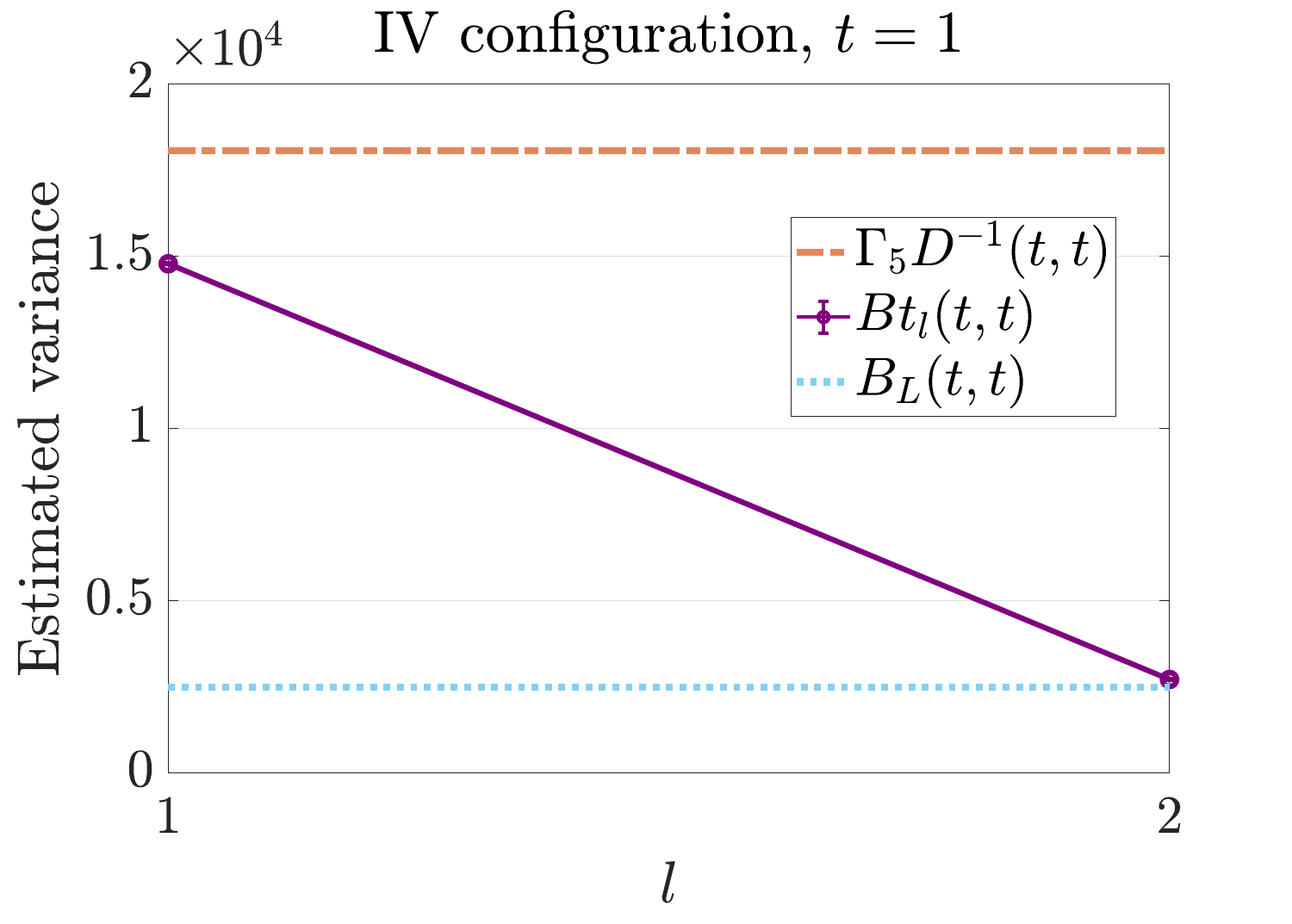}
\includegraphics[scale=0.26]{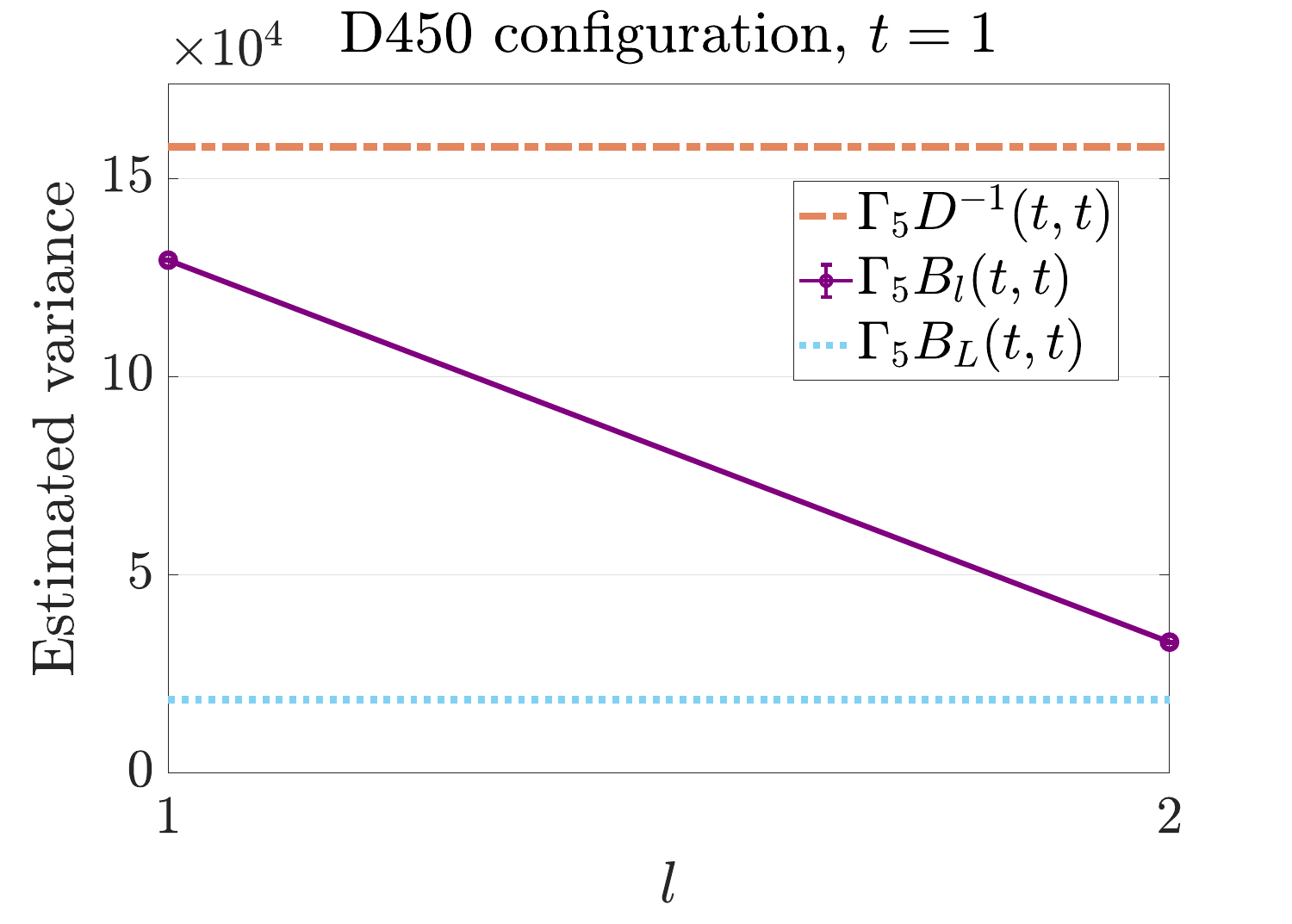}
\caption{\label{fig:disconnected-variances}
Variance reduction for $\tr(\Gamma_5\tr(D^{-1}(t,t))$ with $t=1$ for the IV and D450 configurations. Variances are estimated with a fixed sample size of $N=200$. Shown are the variances for the plain Hutchinson method, the multilevel oblique decomposition $M_l$ ($l=1,2$), and the coarsest-level term $D_L^{-1}$ ($L=3$).}
\end{figure}

\Cref{fig:disconnected-variances} shows a comparison between the plain Hutchinson estimator and the multilevel decomposition from \cref{eq:trace_equality_disconnected}.

We measured the variances of the related operators using three levels with $N_{tv}=28$ test vectors for the construction of the transfer operators and a fixed sample size of $N=200$. This was done for the IV and the D450 ensembles (see \cref{tab:configurations}). The results correspond to the time slice $t=1$. Figure~\ref{fig:disconnected-variances} shows that we achieve a variance reduction at the finest level of slightly less than $18\%$ for both configurations. This is not enough to compensate for the additional cost when compared to just plain Hutchinson. We performed several additional experiments using more levels in the multigrid hierarchy, test vectors which are more precise approximations to small eigenmodes. All of this did not yet result in a method with a lesser cost than plain Hutchinson, and this also holds for the exactly and inexactly deflated Hutchinson variants that we tested.  

\subsection{Discussion}

For the connected operator $G(t)$ in \cref{eq:connected-G}, the variance of the multigrid Multilevel Monte Carlo estimators decreases rapidly with $t$. $G(t)$ involves products of propagator blocks $D^{-1}(t+t',t')$ and $D^{-1}(t',t+t')$ which connect different time slices. Reasoning in terms of eigenmodes of $D$ which decay less rapidly as the eigenvalues increase, at small $t$ the blocks of the inverse still contain strong local contributions, but as $t$ grows the localized parts largely cancel in the spatial trace. What remains at large $t$ is the slowly varying propagation across slices, which is dominated by the low modes of $D$. Since our multilevel decomposition is designed to approximate precisely these slowly varying components in the coarse spaces, it captures the dominant contribution very effectively. As a result, the variance of the multilevel estimators drops quickly with $t$, while the variance of the plain Hutchinson estimator stays approximately constant, leading to the strong reductions observed in ~\cref{fig:connected-traditional}.

For the case of $\tr(D^{-1})$,  contributions confined to a single time slice of the form $D^{-1}(t,t)$, and contributions connecting different time slices of the form $D^{-1}(t+t',t')$, enter. The multilevel decomposition removes mainly the latter, while the former continue to dominate the variance. As a result, the reduction is more modest.

Finally, the disconnected fermion loop case involves only blocks of the form $D^{-1}(t,t)$, and eigenmodes removed or damped by the multilevel decomposition are not dominating the remaining ones on a diagonal time slice in $D^{-1}$.  As a result, the Frobenius norm on these blocks and thus the variance of the disconnected estimators remains essentially unchanged.

\section{Conclusion and Outlook}

We studied variance reduction for three types of problems: the connected pseudoscalar correlator $G(t)$, the  trace of the four-dimensional Dirac operator $\tr(D^{-1})$, and the disconnected fermion loops. For $G(t)$, the multigrid multilevel Monte Carlo estimators achieve strong variance reductions that grow with the time separation $t$, with the orthogonal variand formulation outperforming the oblique one despite involving more terms. For $\tr(D^{-1})$, the overall reductions are more modest and, in our tests, not yet cost-competitive with plain Hutchinson. 

For the disconnected case (time-slice traces built from $D^{-1}(t,t)$ or, more generally, $\tr(\Gamma D^{-1})$ with same-slice blocks), our experiments indicate that the multigrid multilevel Monte Carlo method alone does not provide a meaningful variance reduction. This is consistent with the discussion in \cref{sec:connected_case}: the method targets slowly varying across-slice components, whereas the variance of disconnected loops is dominated by short-range, same-slice contributions. As a result, these problems are not well suited to this method standalone.

Looking ahead, we will explore two complementary directions. First, for the disconnected case, we will investigate techniques that directly target short-range structure, for example probing and related structured-noise strategies, and their combination with multigrid multilevel Monte Carlo. Second, we plan to strengthen the coarse spaces by improving the test vectors used to construct the prolongation and restriction operators (or the deflation space): we will increase their number and compute them to higher accuracy (up to machine precision where feasible) to further reduce the across-slice component.

Preliminary experiments already show promise on the IV ensemble: increasing test-vector accuracy from residuals $\lesssim 10^{-1}$ to $10^{-4}$ and enlarging the number of test vectors from $N_{tv}=28$ to $N_{tv}=64$ yields an additional variance reduction of up to $40\%$ for $\tr(D^{-1})$. For the finest-level term of the connected operator, we observe a $\sim 40\%$ reduction at short distance ($G(0)$) and a reduction by a factor exceeding $2000$ at large separation ($t=24$). We also tested a larger set of 128 test vectors computed to an eigenresidual of $10^{-3}$, observing the same trend toward improved variance reduction. While a more accurate multigrid hierarchy increases setup cost, this overhead can be amortized in practice when many traces are required on the same configuration (for example, multiple time slices or related observables), making the improved hierarchy attractive overall.

\section*{Acknowledgments}
This work is supported by the German Research Foundation (DFG) research unit FOR5269 ”Future methods
for studying conﬁned gluons in QCD”. The computations for the measurements of the variance were carried out on the PLEIADES cluster at the University of Wuppertal, which was supported by the Deutsche Forschungsgemeinschaft (DFG, grant No. INST 218/78-1 FUGG) and the Bundesministerium für Bildung und Forschung (BMBF).
G.R-H. acknowledges financial support from the EoCoE-III project, which has received funding from the European High Performance Computing Joint Undertaking under grant agreement No. 101144014.
The authors gratefully acknowledge the Gauss Centre for Supercomputing e.V. (\url{www.gauss-centre.eu}) for funding this project by providing computing time through the John von Neumann Institute for Computing (NIC) on the GCS JUWELS at Jülich Supercomputing  Centre (JSC).





\bibliographystyle{elsarticle-num}
\bibliography{refs}







\end{document}